*Original Article*

# Design and Architecture for a Centralized, Extensible, and Configurable Scoring Application

Sumit Sanwal

*Independent Researcher, Phoenix, AZ, United States.*

*Corresponding Author : sumitsanwal@gmail.com*

*Abstract - In modern-day organizations, many software applications require critical input to decide the next steps in the application workflow and approval. One of the most important inputs to decide the subsequent course of action is the key performance indicator-based scoring for the entities used in the application. Computing the right score for the entities in the application is a critical step that will drive the subsequent processing and help to decide the next course of action for the entity accurately. Computing the right score is a critical parameter for application processing; deriving the precise and correct score is crucial and pivotal for the application's intended objective; this mandates a very efficient and optimized scoring application in place and is of paramount importance for the success of such applications. We will discuss in this article how to envision and design a generic, extensible scoring engine and a few use cases for scoring with the associated intricacies and complexities to implement the scoring framework.*

## 1. Introduction

Scoring engine is a software application used to compute and assign scores or rankings to various entities based on specific rules from the associated scoring model. These entities could be from different domains like financial, healthcare, e-commerce, etc., ranging from individuals, products, or services to documents, applications, or transactions. The scoring application can be used to compute different kinds of scores. Scoring engines are extensively used in diverse industries and functions; a few examples where they are widely used are in finance, credit scoring, risk evaluation, content streaming, gaming, and more. Organizations in today's world are using scoring applications tailored to their specific needs that are not generic, extensible, and configurable.

This requires high operation costs for any change to scoring logic as well as managing multiple scoring applications within an organization for different kinds of scoring needs, leading to redundancy and maintenance overhead. This article emphasizes and recommend developing a centralizing scoring system which caters to all the scoring requirements within an organization for all kind of disparate scoring needs. The scoring system will be maintained by one central team. It can be designed and developed as a generic, extensible and metadata-driven software which can be integrated with different use cases needing to compute some kind of score. We will discuss a few example use cases and how it can be envisioned as a generic software framework. This will eliminate the need for every application to re-develop and maintain its own scoring functionality from scratch and can be integrated with the centralized scoring system to accomplish any type of scoring needs by leveraging the scoring computation from the central repository.

## 2. Scoring Engine Use Cases

Listed below are a few of the applications, along with the kind of scoring computation to be performed for their respective business needs.

- Computation of Customer score for a loan application. This will be a crucial input for underwriters to adjudicate various kinds of loans, for example, auto loans, home mortgages and credit cards.
- Scoring a business entity for the purpose of approval and review of business loans.

- Scoring customers for e-commerce applications to decide the type of customer and to provide tailored price offerings based on the score to the customer.
- Scoring employees for annual appraisal.
- The scoring model for a balanced scorecard is a tool to help companies identify and improve their internal operations. It provides strategic management performance metrics to help optimize the outcomes.
- A scoring model to compute the score for strategic decisions taken by the company based on the company's strategic indicators.
- The scoring model is to be used by the reward program of the organization to compute the points offered based on the qualifying purchases. These points can be used by the customers to avail themselves of various benefits using the reward points.
- Most of the prominent market segments require scoring computation based on the Key Performance Indicators (KPIs) attributes, which these market segments and domains will use to compute the score:
- In the manufacturing sector, scoring computation will be based on the following KPI attributes: KPI attributes for the financial domain (Gross margin, operating income, return on assets), KPI attributes for the Customer domain (On-time delivery rate, customer satisfaction score, customer retention rate),
- Healthcare Industry: For the healthcare sector, scoring computation will be based on the following KPI attributes: KPI attributes for the financial domain (Operating margin, revenue growth, cost per patient), KPI attributes for the Customer domain (Patient satisfaction score, readmission rate, patient wait time), other KPI attributes used for scoring (Staff turnover rate, hours of staff training, number of new procedures or treatments introduced, Medical error rate, patient safety incidents, average length of stay).
- The retail industry sector will also require score computation based on the following KPI attributes: KPI attributes for the financial domain (Sales growth, gross margin, inventory turnover), KPI attributes for the Customer domain (Customer satisfaction score, customer retention rate, average transaction value), other KPI attributes used for scoring (Employee turnover rate, hours of staff training, number of new products or services introduced, Stockout rate, order fulfillment time, store footfall)

## 3. Need for a Central Scoring Engine

Scoring computation is an intricate and crucial process and involves maintaining distinct kinds of scoring logic. This scoring logic will constitute scoring rules and mappers to compute the score. Again, the application parameters used to compute the score will also vary from application to application and can be managed as a generic reusable metadata-based structure within the scoring solution metadata repository. Users can define the scoring rules based on required scoring attributes and parameters and can configure the scoring rules and associated scoring parameters in the scoring system central repository. Any change to the scoring rules and associated scoring parameters and their values stored in the Mapper configuration can be easily maintained using the user interfaces provided by the central scoring system. This way, the organization will eliminate the need to develop a dedicated scoring system for every application and can reduce a lot of development and infrastructure costs. At the same time, organizations can reduce resourcing costs by having one dedicated team maintaining the scoring system and thus not requiring a dedicated team for every application requiring scoring functionality. This article explains how one central scoring system can cater to all disparate scoring needs, leveraging multiple scoring models for every type of distinct score computation logic along with the associated configurable metadata, rules, and mappers for the given scoring model. Scoring engines are valuable tools that help organizations make data-driven decisions and automate processes that involve evaluating and ranking items or individuals based on specific criteria. They contribute to more efficient, quantifiable, and objective decision-making processes, especially in scenarios involving a large volume of data. This article also recommends a scalable, distributed and cloud-native architecture for the scoring system, which can be easily integrated with any application using Service Oriented Architecture (SOA) or directly using an Application Programming Interface (API) based integration for bulk score processing. The scoring system will have the ability to perform per-record scoring, stream-based scoring, and batch scoring for voluminous datasets. The scoring system is recommended to follow a decoupled architecture where the key components of the scoring engine are completely decoupled as a self-contained service.

All the scoring engine components will be deployed separately in a cloud cluster. They will be integrated with the main component, which will call the other components as per the scoring engine processing needs. The main component will reside in a cloud cluster, so multiple applications can call it, and the load can be balanced using a load balancer to direct the scoring call to the node in the cluster having a lesser load. This can be set up as a scalable, distributed, and cluster-based deployment. The main advantage of this architecture is that it is always up and available to perform the scoring as the processing is broken down into individual decoupled components and deployed in a cluster; even if few of the components are down still, the others will be up and running to perform the processing. This is an entirely decoupled architecture, and the individual components will interact with each other, leveraging the standard communication protocol through API calls and Service-based architecture.

Below is the high-level architecture diagram depicting the entire flow for the scoring engine, along with decoupled services and deployment strategies.

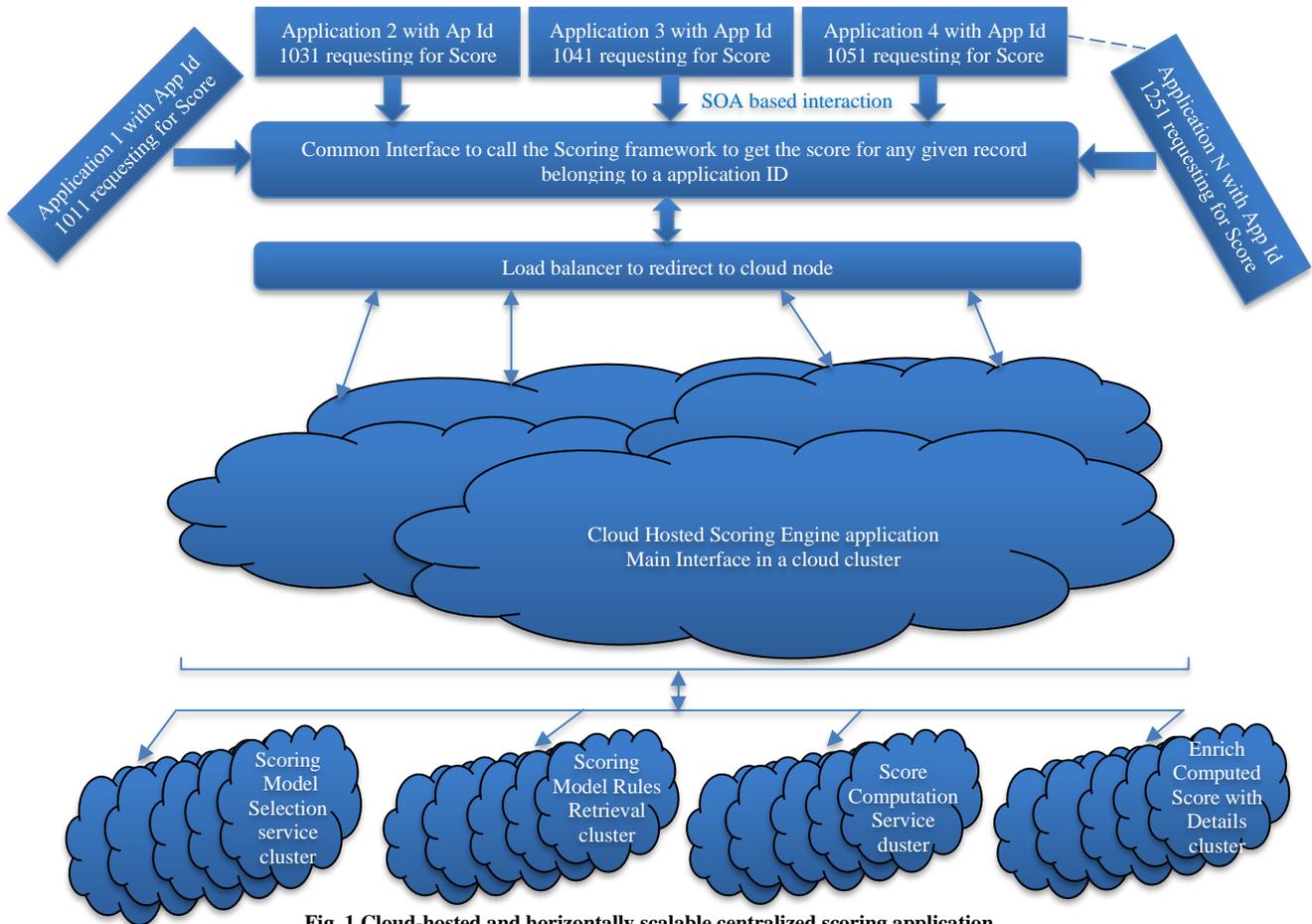

**Fig. 1 Cloud-hosted and horizontally scalable centralized scoring application**

Below is an explanation of the individual components of the scoring system from the above diagram. The entire system is composed of loosely coupled components and will be deployed in a cloud cluster:

### 3.1. Scoring Engine Main Orchestrator Component

The Main Scoring engine component accepts requests to perform scoring. This will be the entry point to trigger the scoring process. This will be deployed as a cloud cluster and will be using a load balancer to redirect the request to the available node as per the load balancing algorithm.

### 3.2. Scoring Model Selection Component

This component will be another self-contained component that will accept the application parameters and determine the most apt scoring model to perform the scoring computation. The scoring model determination algorithm will be Artificial Intelligence (AI) driven. It will be able to detect the best model to be used for scoring based on the application which needs to be scored and the type of scoring to be performed. This can be either system-driven scoring model determination or the application can specify which scoring model to use to skip this step and directly use the specified scoring model to perform the scoring. We will discuss this component in more detail in a subsequent section of this article.

### 3.3. Scoring Mode Rules Retrieval Component

This is the next component of the scoring engine, which will again be deployed in a cloud cluster and used to fetch all the scoring rules and metadata, including the mapping rules for the given scoring model from step 2 above. This whole process will be explained in detail in the subsequent section of this article.

### 3.4. Score Computation Component

This component is a standalone component which will accept the record for which scoring needs to be performed along with the scoring model and all the scoring rules and mappers which will be used to compute the score. This component is the most crucial component in the whole scoring engine and will use the scoring model algorithm to compute the score for the given record based on the scoring rules and rule mapper values. Again, this is recommended to be a standalone component and will be hosted in a cloud cluster and called by the main component based on a load-balancing algorithm to assign the score computation processing to the lesser-used node in the cloud cluster.

### 3.5. Scoring Engine Result Generation Component

Enrich the Computed score with scoring details and return the result to the main scoring component. This component will take the computed score along with the details of how the score was computed and what the influencers are in computing the score. All these details are added to the entity record undergoing score computation, and the enriched record is returned to the main scoring component. This is also deployed in the cloud cluster as a self-contained component.

The central scoring system must be a very efficient and optimized software application that is easily maintainable and allows seamless integration with any application with scoring needs. The scoring system must be generic and extensible, which can quickly adapt to any changing scoring logic and easily incorporate any new processing pattern. The system should provide a simplified interface to maintain the scoring logic. The system should have the ability to respond to any change in scoring logic or mapper rules in a very agile manner, and any frequent scoring rules or logic change can be incorporated into the system with a quick turnaround time, thus providing the ability to roll out any changes to score computation logic swiftly.

## 4. Key Characteristics and Functionalities of a Scoring Engine

### 4.1. Scoring Algorithm

The scoring engine employs a specific algorithm or model to process the input data and generate scores. The scoring model can have different kinds of computation logic algorithm. The few most frequently used scoring algorithms are rule-based logic, weighted average algorithm, statistical score computation algorithm, machine learning-based which self-improvises and optimizes itself, and Natural Language Processor (NLP) based scoring model to evaluate essays and assign scores. The scoring model algorithm can be a hybrid model, too, using a combination of these approaches.

### 4.2. Rules and Criteria

A scoring engine operates based on a scoring model, which is nothing but a score computation algorithm explained in the above point. Every scoring model will have its associated predefined rules and criteria. These rules are used to find the matching values and weights for the given scoring model as per the input data from the record which needs to be scored. These rules can be as simple as a mapper rule to a complex rule like an expression-based scoring condition to find the matching values from the scoring model. These rules and criteria are used to evaluate the input data and determine the resultant scores using the scoring model algorithm.

### 4.3. Data Input

Scoring engines take data as input from the application, which needs to perform scoring for a given entity; the application provides the entity details to the scoring engine as input. The input traits of the entity can be in the form of a structured input like a database record, or for SOA-based approach; it can be in the form of any format like a JSON or XML or a properties file or in the form of unstructured traits like an image or text data. The type and format of input data fed to the scoring engine will depend on the type of scoring to be performed, the scoring model to be used, and what kind of input that scoring model will require.

### 4.4. Integration

Scoring engines operate in different modes, which include silo mode, where it must score a specific entity, or it can score a bulk of entities. The scoring engine provides complete automation and seamless integration to process scoring for the given entities quickly and efficiently using SOA-based integration, thus providing near real-time scores or also as backend processing using bulk mode, which runs periodically as per the defined frequency and updates the score for all the given bulk entities in the application using API based integration where the given application calls the scoring engine services integrated into the application logic and uses it to perform score computation as part of the bulk processing.

### 4.5. Customization

In many cases, scoring engines should have features like extensibility, which is needed to suit the needs of a business or application. This may involve incorporating a new score computation algorithm to compute the score, or it may also mean fine-tuning an existing scoring algorithm. The underlying scoring engine should have the ability to add a new score computation pattern or to change an existing pattern, so in a nutshell, it should be highly customizable.

### 4.6. Data Output

The output of a scoring engine is typically a score or rating assigned to each input entity, reflecting its relative quality, risk level, or any other relevant metric.
Listed below are a few widely used use cases requiring specific type of score computation:

### 4.7. Credit Scoring

Banks and financial establishments will require the creditworthiness of individuals and businesses. They need the credit score of the applicant to adjudicate the credit applications.

### 4.8. Risk Assessment

Scoring engines are used in various domains like insurance and investments to assess the risk associated with prospective customers, and the score is a quantified outcome from the scoring engine which helps evaluate and mitigate risks.

### 4.9. Content Filtering

Scoring engines are used in the content filtering world where every content is scored, and based on the score outcome, the decision is made to include or mark the content as spam; it also helps in sentiment analysis to determine the significance and aptness of content.

### 4.10. Gaming

The scoring engine is widely used in gaming applications to determine various kinds of scores, including players' scores and relative rankings, as well as different levels of gaming to determine complexity scores.

### 4.11. Ranking and Recommendations

Scoring engines are also used in various kinds of ranking use cases like appraisal ratings, product ratings, service ratings based on customer feedback or content ratings based on user preferences and historical data.

For computing the right score for an entity in a specific use case, the first step is to identify the Key Performance Indicators (KPIs). Key Performance Indicators (KPIs) are a set of critical data elements for the given entity in the use-case to assess the quality and performance of the entity. KPI attributes specifically help determine the use-case's performance and guidance, especially compared to those of other peers within the same domain.

Listed below are a few domains which require the main entities used in the application to be scored leveraging all the identified KPI attributes. As a next step, the right scoring model will be fetched. Based on the fetched scoring model, scoring rules and rule mapper matrix will be fetched from the scoring metadata. Next, the scoring engine will use all this metadata information to compute the precise score for the application. This whole framework can be developed as a centralized, generic, and configurable application that can be leveraged by multiple applications requiring any scoring to be performed as part of the application business requirements. The computed score from the scoring engine will be used by applications to be used for the subsequent processing and to achieve the business objective.

## 5. Generic Scoring Engine Design

Designing a generic scoring engine involves creating a flexible system to evaluate different entities based on various criteria. Below is a high-level design of a generic reusable scoring engine that various applications can use to perform the scoring.

The applications can interact with the scoring engine using a common interface, shown below as ScoringEngineService, for illustration purposes. The application, while calling the scoring engine service, will pass the application ID, the Scoring model to be used to perform the scoring (this is optional; in case not provided, the scoring engine will determine the scoring model to be used), a record for which the scoring to be performed (Map of the record with the record attributes and the values) and the last argument is the key performance indicators for the application. The scoring engine should accept input data representing the entities to be scored. This data could be in many formats, such as structured data (e.g., JSON, XML, CSV, properties file) or unstructured data (e.g., images, icons, Texts).

Based on the above-listed parameters, the scoring engine will perform the scoring for the given record from the application. The whole scoring is divided into a four-step process, as depicted and described below.

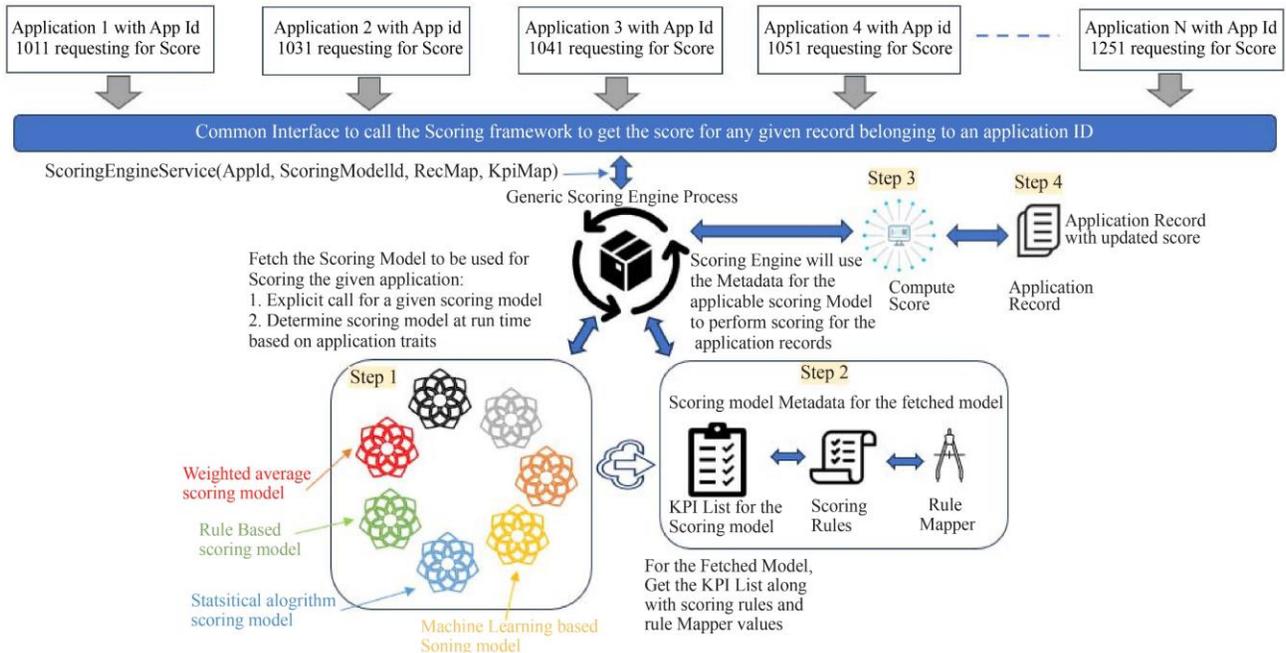

**Fig. 2 Centralized Scoring Engine end-to-end flow computing scoring for multiple applications using different kinds of scoring models**

Step 1: Scoring model selection for score computation: in this step, the scoring model to be used to perform the scoring will be determined. The scoring engine supports various models to compute the score. The scoring model algorithm can be a weighted average score of the key performance indicators, or it can use the historical trending of the customer to compute the score based on how the applicant has done in recent years. Every scoring model requires a different kind of computation algorithm, and based on the type of application and score to be computed, the scoring model selection process will use the input parameters and the type of entity to be scored to determine the most apt model to compute the score. Application requesting a score will pass the scoring model identifier in case it has a pre-decided scoring model to be used for score computation, or the application will leave it to the scoring model selection component to find the best model to be used to compute the score. Suppose the application does not request a specific scoring model to be used for scoring. In that case, the scoring engine will assess the type of application based on the application ID, the application traits, and the kind of scoring it requires. Next, the scoring engine will use the application characteristics and the key performance indicators passed as another set of arguments to the scoring service; the scoring engine will determine the most apt scoring model for the application scoring. The Scoring engine has a scoring model determination process to find the best-fitting model based on the KPIs and the type of application which requires the scoring. Application ID will help to determine the type of application and what kind of scoring is requested by the application.

**Custom Scoring Model used by Scoring Engine:** End user and business analysts managing the scoring models to be used for score computation must define the scoring rules in such a manner that it clearly flags out outlier entities who are high risk for the business and can lead to negative return on investment for the business. The typical business product represents a way for organizations to be profitable and can bring in a reasonable return from the business. The scoring model is a critical tool for assessing prospective customers to conduct business and remain profitable.

The risk of loss in business, also known as dealing with high-risk customers as well as non-profitable entities, may preclude a business from being repaid and is the first concern of the organization. Thus, the associated risk with a particular business transaction will be the basis for the pricing offered to the customer. Also, it will limit the exposure to high-risk customer profiles.

The Customer Scoring Model will use the business entity information to calculate the associated risk. The scoring model will characterize the entity and its characteristics using key entity characteristics into a set of parameters, which will be used to assess the entity in quantitative terms and will also become the base for performing business and deciding and formulating the transaction term with the entity by the transacting organization.

Parameters are the economic factors against which the scoring model is assessed. Each parameter has variables on which score values are entered. The score values denote the preference of one element over another under a particular parameter. Also, each parameter is given a weight denoting its importance. Not all parameters need to have the same importance or weightage and to differentiate their importance, each parameter is given a weightage.

Step 2: Fetch the scoring rules and mapper values for the given scoring model. Once the scoring model is finalized, the next step is to retrieve the scoring rules and the rule mapper values for the given scoring model. This will require the scoring engine to retrieve the scoring metadata from the scoring config repository based on the identified scoring model to be used. Every scoring model might be using a different kind of scoring metadata. So, basis the scoring model, the respective scoring metadata will be retrieved. Retrieving the right set of scoring rules for the given scoring model that will be used to evaluate the entity is a complex process. Each scoring rule should have an identifier, rule name, weight, and a scoring method. The weight represents the significance of the criterion in the overall score, and the scoring method defines how to calculate scores for that criterion. All the pertinent scoring rules, along with scoring model mapper values required to perform the scoring for the given record from the application, will be retrieved.

Step 3: Compute score for the given input entity based on the selected scoring model: In this step, the score is computed for the given record passed by the application. The record attribute values are fetched from the record Map passed as the third argument and used to run the scoring rules and compute the score. The computation process will use the record values against the applicable rules and determine the score based on the rule mapper by leveraging the KPI elements used in the scoring logic.

The score computation process will trigger the scoring model process that can apply the scoring rules to the input data. The rule engine should be able to handle both simple rules (e.g., if-else conditions) and complex rules (e.g., rule-based expressions). Every scoring model will have its scoring algorithms to calculate scores based on different data types and criteria. For example, you may need algorithms for numerical data, text data, image data, etc. The scoring model logic will be configured by the scoring team to customize the scoring model and algorithms. This could be achieved through configuration files, APIs, or user interfaces.

Remember that the specific design details of a scoring engine will depend on the requirements of your use case. By creating a flexible and modular design, you can easily extend the scoring engine to handle different types of entities and criteria in the future.

I will explain this with a few other detailed scenarios in the subsequent section once the score is computed based on the scoring model logic. The same is passed by the engine to the next step.

Step 4: Score assignment and returning the result to the application: in this final step of the process, the computed score is enriched in the record, and the updated record Map is returned by the scoring engine back to the application. In a few scenarios, the scoring engine will also provide more insight into the KPI elements, how the score was computed and which KPI element impacted how to compute the score. This will help the application to get more detail behind the scoring and how the record KPI elements played their part in determining the score. The returned record Map contains all the details and will be used by the calling application to use this insight in the subsequent application processing.

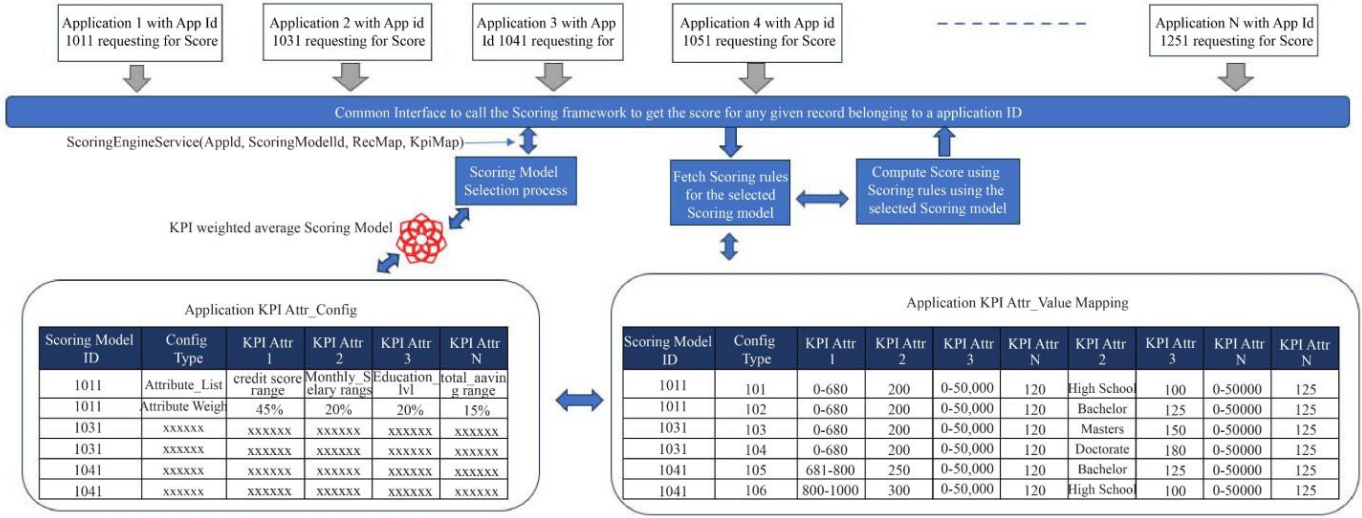

**Fig. 3 Scoring Application high-level flow computing score using a weighted average scoring model**

## 6. Score Computation Use-Case

In this section, we will discuss one of the scoring models which is very extensively used and will demonstrate how to compute the score for the given use case using the given scoring model.

In the above diagram, to compute the score, the following detail processing will be followed:
1. The application initiates a call to perform a record score for the given application ID and record.

2. Scoring engine common interface was invoked with the required parameters, including the application Id, scoring model ID, record object, and KPI attribute list.

3. In this case, the application specifically asked to use the scoring model 142, which is a model using a weighted average of the key performance indicator attributes to compute the score. In case the scoring model is not explicitly specified in the call to the scoring framework, in those scenarios, the scoring framework will determine the scoring model based on the application parameters, type of scoring to be performed, and the KPIs to be used to perform the scoring from the KPI list parameter. All these arguments are considered, and Scoring model selection criteria are triggered to find the matching scoring model based on the arguments. Scoring model selection criteria is an intelligent algorithm that finds the best matching scoring model based on the input parameters.

4. As a next step, the scoring model rules metadata is fetched, which is as follows:

a. Weighted average rule Metadata: The indicators used to compute the score and the respective weight for the indicators. This is depicted in the above diagram's first metadata table for the scoring model.

b. Mapper rules for Score computation: Scoring rules will constitute all combinations of indicator values/ranges and the respective score for every indicator in the given combination. This is depicted in the second metadata table in the above diagram.

5. In this step the score is computed based on the retrieved scoring model rules and the record for which the scoring will be performed. Given below the flow diagram to compute the score for the given weighted average scoring model highlighted in red in below diagram. Also, explained every step for score computation as part of the below points:

a. Find the values of all the KPI attributes from the given record for which the scoring needs to be performed. In the below example, the KPI attributes for the given applicant record with ID '104532' to be used in scoring are as listed below:

    i. Credit Score = 790
    ii. Monthly Salary=12000

   iii.   Highest Education Level= Bachelor
   iv.   Total Bank Saving=30000

b.  For the given values, find the matching record from the mapping rules. Pick the individual ratings for every KPI attribute based on the matching record. In the below snapshot, the applicant entity to be scored is matched against the scoring model rule and for the identified scoring mode, the matching rule (Rule Mapping ID 105) is highlighted in green below based on the following matching criteria:

   i.   Credit Score: 790 in the range <600 − 800 > &&
   ii.   Monthly Salary :12000 in the range <0 - 50000 > &&
   iii.   Highest Education Level:  Bachelor = Bachelor &&
   iv.   Total Bank Saving: 30000 in the range <0 - 50000 >

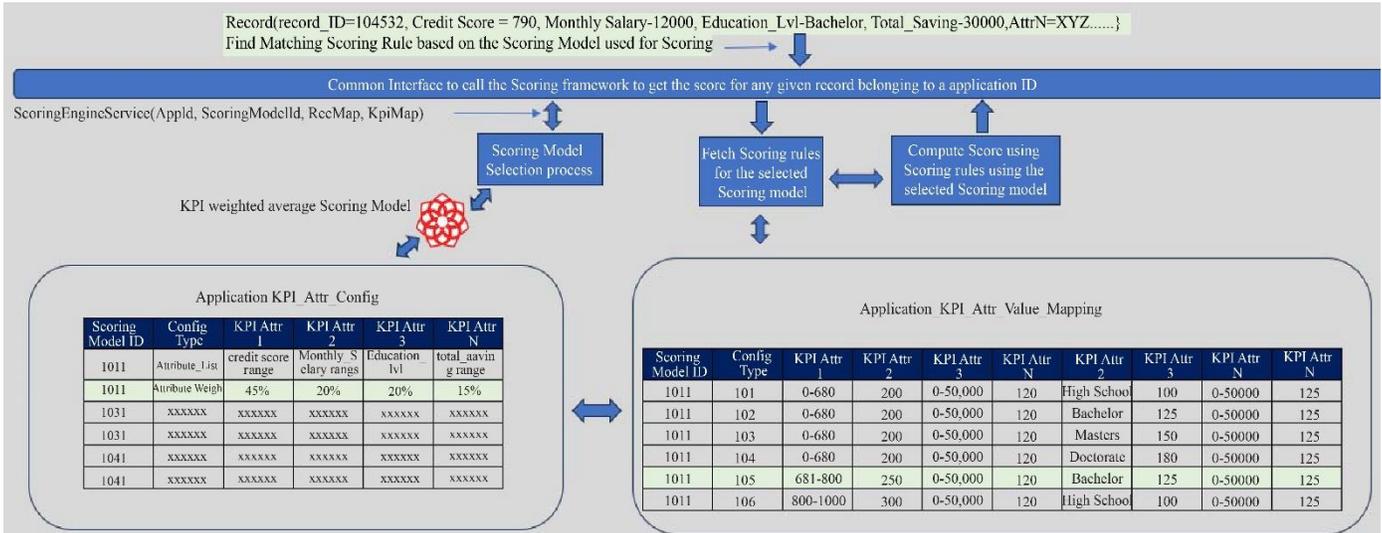

**Fig. 4 Scoring application computing score for an entity using a weighted average scoring model**

c.  Next, compute the weighted average based on the first metadata table using the weight for every KPI attribute for the given Scoring model (Scoring Model 1011). The weighted average computed is the score for the given record. Score Computation below is for a given record with matching scoring rules (Rule Mapping ID 105) and using the weighted average computation rules for the scoring model 1011 to compute the score.

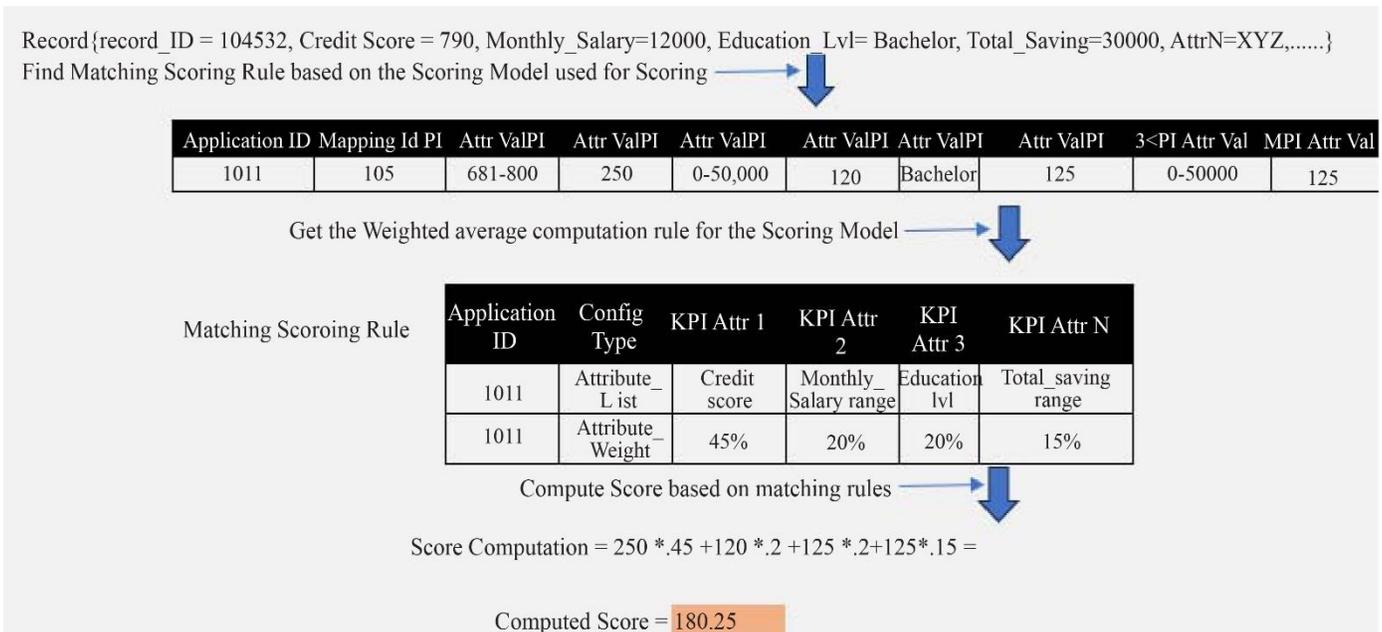

**Fig. 5 Score computation flow for an entity using the weighted average scoring model**

d.  Next step is to update the score in the record itself and return the enriched record as the return value to the calling application.

Record{record_ID = 104532 , Credit_Score = 790, Monthly_Salary=12000,Education_Lvl= Bachelor, Total_Saving=30000, Computed_Score = 180.25 , AttrN=XYZ,......}

**Fig. 6 Snapshot of computed score leveraging a weighted average scoring model**

The system should produce internal custom scores based on business-defined scoring rules. The system should allow the user, based on his / her authority level, to configure the scoring rules for the scoring model for score computation. In addition, the system should also provide the ability to add/modify/delete scoring rules to keep the scoring model relevant to adapt to changing dynamics influenced by internal and external factors impacting the score computation algorithm.

The scoring engine should have the following core capabilities:

1. The Scoring engine should have the ability to automatically attach a scoring model with the prospective entity to be scored. The system should be able to associate the scoring model manually or auto-select using a scoring model selection feature.
2. Scoring framework should allow defining any number of scoring models. Business users should be able to add criteria/rules for the Scoring model using the user interface.
3. The scoring framework should provide the flexibility to the business to use data elements (once captured) as attributes to be used by the selection feature to select a scoring model.
4. The scoring engine should display both the score and the underlying attributes and calculated values used to derive that score as a return output from the scoring engine.

## 7. Custom Scoring Model used by Scoring Engine

End users and business analysts managing the scoring models to be used for score computation must define the scoring rules in such a manner that it clearly flags out outlier entities who are high risk for the business and can lead to negative return on investment for the business. The typical business product represents a way for organizations to be profitable and can bring in a reasonable return from the business. The scoring model is a critical tool for assessing prospective customers to conduct business and remain profitable.

The risk of loss in business, also known as dealing with high-risk customers as well as non-profitable entities, may preclude a business from being repaid and is the first concern of the organization. Thus, the associated risk with a particular business transaction will be the basis for the pricing offered to the customer. Also, it will limit the exposure to high-risk customer profiles.

The Customer Scoring Model will use the business entity information to calculate the associated risk. The scoring model will characterize the entity and its characteristics using key entity characteristics into a set of parameters which will be used to assess the entity in quantitative terms and will also become the base for performing business and deciding and formulating the transaction term with the entity by the transacting organization.

Parameters are the economic factors against which the scoring model is assessed. Each parameter has variables on which score values are entered. The score values denote the preference of one element over another under a particular parameter. Also, each parameter is given a weight denoting its importance. Not all parameters need to have the same importance or weightage and to differentiate their importance, each parameter is given a weightage.

## 8. Scoring Model Definition and set up for the Lending Application

In this section, we will do an in-depth assessment of one of the most prominent use cases using scoring. We will focus on the use case for the mortgage domain, and here we are lending to customers for various kinds of loans, including home loans, auto loans and personal loans.

To evaluate the potential buyers' credit merit, we will design and model a broad scoring model to evaluate them on all the critical parameters. This will be a specific scoring model based on the key performance indicator attributes that assess the customer risk profile to establish the probability that the buyer will repay loans. Explained below is the scoring model which will be used for calculating the score for the entities from the lending application.

The scoring model for the mortgage use case calculates scores to assess whether a loan should be approved or not. Using the scoring model, banks adjudge mortgage requests based on the score from the scoring engine. The Scoring engine computes the score for the main borrower and any co-borrower based on the selected characteristics of the potential borrower.

The scoring model for loan application is modelled based on the 5 Cs of credit, i.e., Credit, Character, Collateral, Conditions and Capacity. These 5 Cs are divided into several factors by which the borrower and the co-borrowers are measured, and a final score is calculated. This final score helps the adjudicators decide the loan approval eligibility of the borrower applicant, and based on the customer's risk profile, interest rates and pricing are defined for the loan. The pricing can also be determined using a pricing engine, which again will set the interest rates and pricing terms based on the borrower's creditworthiness.

The following diagram depicts the scoring model key scoring parameters setup for the calculation of the final score:

In the below model, the marks and the average weighted score for every parameter are computed based on the parameter rules defined in the scoring model. In the table below, one applicant/Co-Applicant is scored based on the parameter value for every parameter, and computing the score is using the parameter rule from the scoring model as depicted for one applicant and co-applicant record where "XXXX" represent the score for the applicant and co-applicant.

| S. No. | Key Scoring Parameters | Applicant param value / Co-Applicant param value | ATTRIBUTE WEIGHTAGE | Primary Applicant | | Co-Applicant | | Aggregate Weighted Score |
|---|---|---|---|---|---|---|---|---|
| | | | | Marks | % Weightage | Marks | % Weightage | |
| I | Applicant Age | 35 / 44 | 10 | XXXX | 65 | XXXX | 35 | XXXXX |
| ii | Applicant Employment Type | Business / Accountant | 15 | XXXX | 60 | XXXX | 40 | XXXXX |
| iii | Applicant Monthly Income | 10,000 / 12,000 | 15 | XXXX | 50 | XXXX | 50 | XXXXX |
| iv | Applicant Credit Score | 790 / 805 | 20 | XXXX | 50 | XXXX | 50 | XXXXX |
| v | Outstanding Judgments | None/one | 10 | XXXX | 50 | XXXX | 50 | XXXXX |
| vi | Years of School | 16 / 18 | 10 | XXXX | 65 | XXXX | 35 | XXXXX |
| vii | Intended use of Property | Investment / Investment | 10 | XXXX | 60 | XXXX | 40 | XXXXX |
| ix | Outstanding Judgments | 0 / 1 | 10 | XXXX | 60 | XXXX | 40 | XXXXX |

**Fig. 7 Cloud-hosted and horizontally scalable centralized scoring application**

## 9. Conclusion

Considering all the key features discussed in this article, organizations can design and develop a centralized, extensible, and configurable scoring application and can use it to cater to all the scoring requirements for the organization. All the core principles explained in this article will help organizations to design and develop an efficient scoring framework optimally. The main concept emphasized in this article is to design and develop a maintainable and loosely coupled framework which can be extended as needed to add more score computation patterns in the form of a scoring model. The scoring framework is emphasized to be a central and common application for the organization. It should support score computation for different use-cases comprising multiple domains and products like for mortgage use case, e-commerce use-case as explained in this article and any other use-case that can be supported by the same central scoring application. This article explains that the scoring application can be developed using a modern technology stack, providing robust scalability and distributed computation capabilities. This can be hosted in the cloud and designed as a market-in-a-box product deployed in any geography with minimal effort. Also, the scoring application can provide APIs to process scores for the downstream applications. Hope this provides a few different perspectives to the users to design modern scoring applications efficiently.

## Conflicts of Interest

* Disclosure of potential conflicts of interest:
  On behalf of all authors, the corresponding author states that there is no conflict of interest.
* Research involving human participants and/or animals:
  This article does not contain any studies with human participants or animals performed by any of the authors.
* Informed consent:
  No individual participants were included in the study, so no consent was needed.

## References

[1] Pierre Veyrat, Balanced Scorecard Examples and their Application in Business. [Online]. Available: https://www.heflo.com/blog/balanced-scorecard/balanced-scorecard-examples/

[2] Scott Wood, What is an Automated Scoring Engine? | Augmented Intelligence, 2020. [Online]. Available: https://medium.com/actnext-navigator/what-is-an-automated-scoring-engine-augmented-intelligence-7914f39cf380

[3] Paul Wilson, How an AI Scoring Engine will Drive Customer Engagement, 2018. [Online]. Available: https://www.linkedin.com/pulse/how-ai-scoring-engine-drive-customer-engagement-paul-wilson

[4] Paul Nelson, Search Engine Scoring: A New Age of Search Engine Relevance. [Online]. Available: https://www.accenture.com/us-en/blogs/search-and-content-analytics-blog/search-engine-scoring-relevance

[5] Diana Gorea, "Knowledge as a Service. An Online Scoring Engine Architecture," *The Third International Multi-Conference on Computing in the Global Information Technology*, pp. 1-6, 2008. [CrossRef] [Google Scholar] [Publisher Link]

[6] Frederic Fillouxs, Scoring Stories to Make Better Recommendation Engines for News, 2017. [Online]. Available: https://mondaynote.com/scoring-stories-to-make-better-recommendation-engines-c3c73a596893

[7] Kevin Gunning, and Kathy Rowan, "Outcome Data and Scoring Systems," *BMJ*, vol. 319, 1999. [CrossRef] [Google Scholar] [Publisher Link]

[8] Diana Gorea, "Dynamically Integrating Knowledge in Applications. An Online Scoring Engine Architecture," *9th International Conference on Development and Application Systems*, pp. 216-221, 2008. [Google Scholar] [Publisher Link]